# Veritas-RPM: Provenance-Guided Multi-Agent False Positive Suppression for Remote Patient Monitoring

*Architecture Design and Simulation Validation*

**Aswini Misro[1], Vikash Sharma[2], Shreyank N. Gowda[3]**

[1] YouDiagnose Limited / NHS, London, UK, [2] YouDiagnose Limited, London, UK, [3] School of Computer Science, University of Nottingham, Nottingham, UK

*Correspondence: aswini@youdiagnose.com*

Code repository: https://zenodo.org/records/19136793 · GitHub: https://github.com/justbetter21/veritas-rpm

## Abstract

**Background:** Remote Patient Monitoring (RPM) deployments across the NHS generate false positive alert rates of 74-99%, producing alarm fatigue that degrades clinical response and patient safety. Existing RPM systems predominantly rely on threshold-based alerting, with limited integration of contextual factors such as activity, device integrity, and patient-specific baselines.

**Methods:** We present Veritas-RPM, a provenance-guided multi-agent architecture comprising five processing layers: VeritasAgent (ground-truth assembly), SentinelLayer (anomaly detection), DirectorAgent (specialist routing), six domain Specialist Agents, and MetaSentinelAgent (conflict resolution and final decision). We construct a 98-case synthetic taxonomy of false-positive scenarios derived from documented RPM patterns. Synthetic patient epochs (n = 530) were generated directly from taxonomy parameters and processed through the pipeline. Ground-truth labels are known for all cases. Performance is reported as True Suppression Rate (TSR), False Escalation Rate (FER), and Indeterminate Rate (INDR).

**Results:** Overall TSR was 83.7% (82/98 cases). FER was 16.3% (16/98). INDR was 0% - no case was left unresolved. All evaluated cases across five of the six specialist agent domains achieved 100% TSR (noting that two domains had small case volumes: BradycardiaAgent n=2, NocturnalAgent n=3, with wide confidence intervals). The remaining performance categories - TachycardiaAgent, MetaSentinelAgent conflict resolution, and two multi-signal overlap subcategories, are reported separately as they involve routing interactions between agents rather than single-domain decisions. Note that the 100% TSR for ProbeIntegrityAgent refers to single-domain cases (n=23); multi-domain routing cases involving probe integrity are reported separately and show lower performance: ProbeIntegrityAgent (n=23), ActivityIntegrityAgent (n=8), COPDAgent (n=13), BradycardiaAgent (n=2), and NocturnalAgent (n=3). Failures concentrated in MetaSentinelAgent conflict resolution (TSR 70%, FER 30%, n=30) and multi-signal overlap cases (ProbeIntegrity + condition conflicts: TSR 0%, n=3). Failure mode analysis identified system_flag device status (n=7) and threshold-marginal states (n=1) as the primary drivers of MetaSentinel false escalations.

**Conclusions:** Provenance-guided multi-agent routing achieves high false positive suppression across single-domain alert types in simulation. Failure modes are architecturally localised and interpretable, identifying a precise target for real-world validation. This study demonstrates architectural feasibility rather than clinical performance. Phase 1 prospective validation on 200 curated NHS RPM cases with dual-specialist conformity assessment is planned.

**Keywords:** remote patient monitoring; multi-agent systems; false positive suppression; alarm fatigue; NHS; clinical decision support; provenance-guided AI; agentic AI

# 1. Introduction

## 1.1 The false positive problem in NHS remote patient monitoring

Remote Patient Monitoring (RPM) has become a central pillar of NHS virtual care strategy, with the Long Term Plan's 40,000 virtual ward bed target driving rapid expansion of home-based physiological monitoring across acute and community settings. Several commercial platforms now provide continuous vital sign monitoring to large patient populations across NHS trusts. Despite this scale, alarm fatigue generated by false positive alerts remains a well-documented and operationally significant problem in both inpatient and remote monitoring contexts (Sendelbach & Funk, 2013; Drew et al., 2014).

Published evidence consistently shows that a substantial proportion of physiological monitor alarms are non-actionable, with reported rates ranging from 74% to 99% depending on monitoring context and device configuration (Cvach, 2012; Drew et al., 2014). In remote monitoring settings, false positive alerts constitute a large fraction of transmitted events, with studies reporting rates such as 59.7% in implantable device monitoring cohorts (Covino & Russo, 2024). Similar trends have been observed in wearable and telehealth systems, where motion artefact, signal degradation, and patient-specific physiological variation contribute significantly to non-actionable alerts (Charlton & Marozas, 2022; Elgendi, 2012).

The consequences of high false positive rates are well established. When alert streams contain a high proportion of non-actionable notifications, clinician response times to genuine events increase, override rates rise, and cognitive load associated with alert processing displaces higher-value clinical tasks (Ancker et al., 2017). Studies of clinical decision support systems have reported that only a small fraction of alerts are clinically appropriate, with override rates exceeding 90% in some settings (Ancker et al., 2017). Similar findings have been reported across broader clinical monitoring literature, with non-adherence rates ranging from 49% to 96% (Simpson & Lyndon, 2019). In such environments, excessive alerting degrades the signal-to-noise ratio required for timely and effective clinical escalation.

## 1.2 Current NHS RPM platforms and the suppression gap

Several commercially deployed UK RPM platforms now operate at scale within NHS trusts. Whzan Digital Health's BluetoothKit collects vital signs and generates alerts via NEWS2 scoring logic. Luscii (Graphnet Healthcare) applies configurable threshold rules for SpO2, heart rate, blood pressure, and temperature. Doccla provides continuous monitoring with a clinician-facing dashboard. Huma operates a cloud platform with configurable alert thresholds and patient-reported outcome integration.

A common architectural characteristic across these platforms is that alert generation is primarily based on threshold exceedance at the level of individual physiological parameters. While threshold-based systems are effective for detecting gross deviations, they are known to be sensitive to artefacts arising from motion, sensor placement, and signal quality issues (Clifford et al., 2012; Charlton & Marozas, 2022). In practice, this can result in alerts being generated without sufficient consideration of contextual factors such as patient activity, device integrity, or individual clinical baselines.

More recent work in clinical monitoring and wearable sensing has highlighted the importance of incorporating contextual information, including activity state, signal quality indices, and patient-specific baselines, to improve alert reliability (Plesinger et al., 2016). However, these approaches are typically implemented as enhancements to individual signal-processing or classification components, rather than as an explicit system-level suppression layer integrating multiple contextual dimensions.

Existing RPM platforms therefore primarily rely on threshold-based alerting with limited integration of multi-source contextual reasoning. This gap motivates the need for architectures that explicitly incorporate provenance, context, and domain-specific reasoning into the alert suppression process.

### 1.3 Contribution of this paper

We present Veritas-RPM: a provenance-guided multi-agent reference architecture designed as a suppression and classification layer that can operate downstream of any existing NHS RPM alert-generating system. The architecture's primary innovation is a data provenance layer that structurally separates verified ground-truth inputs from LLM inference before clinical reasoning begins, enabling downstream specialist agents to distinguish physiological signal artefacts from genuine deterioration events without relying on post-hoc correction.

This paper describes the system architecture, study design, and results of a simulation-based validation on a 98-case false-positive taxonomy. The study is positioned as a pre-clinical evaluation consistent with FDA and MHRA guidance on simulation methodology for medical device development. A Python reference implementation has been published on Zenodo (DOI: 10.5281/zenodo.19136793) and GitHub (https://github.com/justbetter21/veritas-rpm).

## 2. Related Work

### 2.1 Alarm fatigue and false positive burden in clinical monitoring

Alarm fatigue is a well-documented challenge in both inpatient and remote monitoring settings, with non-actionable alert rates frequently exceeding 70% across physiological monitoring systems. Foundational studies in clinical environments have shown that high alarm volumes lead to increased override rates, delayed responses, and reduced clinician trust in monitoring systems (Sendelbach & Funk, 2013; Drew et al., 2014). Similar patterns have been observed in clinical decision support systems, where inappropriate alerting contributes to override rates exceeding 90% in some contexts (Ancker et al., 2017).

A range of approaches have been proposed to reduce false positive alerts. Signal-processing methods attempt to filter noise and artefacts at the waveform level, while machine learning approaches learn discriminative patterns from labelled alert streams. For example, anomaly detection frameworks based on autoencoders and statistical modelling have been applied to physiological signals to reduce false alarms (Zong et al., 2018), and community benchmarks such as the PhysioNet Challenge have focused on algorithmic reduction of ICU false alarms. Personalised thresholding approaches further attempt to adapt alert criteria to individual patient baselines.

Despite these advances, most existing approaches operate on individual signal streams in isolation. In contrast, real-world RPM alerts are often driven by interactions between multiple contextual factors, including motion artefact, probe displacement, patient activity, and underlying clinical conditions. Prior work in wearable monitoring has demonstrated that photoplethysmography (PPG)-based measurements are particularly sensitive to motion artefacts and signal quality degradation (Charlton & Marozas, 2022; Elgendi, 2012), reinforcing the need for context-aware interpretation beyond single-parameter thresholding.

### 2.2 Multi-agent systems in clinical AI

Recent work in multi-agent AI has explored decomposing complex decision-making tasks into coordinated sub-agents. In clinical contexts, such approaches have been applied to diagnostic reasoning and question answering, where multiple specialised agents contribute to a final decision. For example, ConfAgents (Zhao et al., 2025) introduces a conformal prediction-based routing mechanism that selectively escalates uncertain cases to specialist models, improving computational efficiency.

Parallel developments in large language model (LLM) systems have demonstrated the potential of agent-based architectures for structured reasoning, tool use, and task decomposition. Frameworks such as AutoGen and related multi-agent orchestration systems highlight the benefits of modular reasoning pipelines, though their application in continuous physiological monitoring remains limited.

While prior work focuses primarily on diagnostic reasoning tasks or static query settings, Veritas-RPM addresses a distinct problem: continuous monitoring and suppression of non-actionable alerts in physiological data streams. Rather than optimising accuracy on benchmark datasets, the objective is to reduce false escalations in a high-frequency, safety-critical setting. The proposed architecture differs in explicitly structuring decision-making across multiple domain-specific agents, coordinated through a deterministic aggregation layer informed by data provenance.

## 2.3 Simulation validation in medical device development

Simulation-based validation is an established methodology for pre-clinical evaluation of medical device software, supported by regulatory guidance from both the FDA and MHRA. Synthetic data generation enables controlled evaluation against known ground-truth labels, which is difficult to achieve with real-world clinical data in early-stage development.

Recent work in healthcare simulation has explored the use of synthetic patient data and digital twin models to evaluate clinical decision systems under controlled conditions. Platforms such as Synthea have demonstrated the feasibility of generating realistic, privacy-preserving synthetic health records for benchmarking and validation. In the context of physiological monitoring, simulation enables systematic exploration of edge cases, including rare artefact combinations and threshold-marginal scenarios.

However, simulation-based evaluation introduces inherent limitations, including distributional alignment between the synthetic dataset and the assumptions encoded in the system. As such, simulation is best viewed as a tool for architectural validation and failure mode analysis, rather than as evidence of real-world clinical performance.

## 3. System Architecture

### 3.1 Architecture overview

The Veritas-RPM pipeline comprises five sequential processing layers, each consuming a defined data object from the layer above. Context flags embedded at the input stage flow as provenance tags through the entire pipeline, enabling each downstream component to access the full measurement context without requiring repeated re-query of raw data sources.

| Table 1. Veritas-RPM five-layer pipeline: components, data objects, and primary functions. | | | | |
| --- | --- | --- | --- | --- |
| Layer | Component | Input | Output | Primary function |
| 1 | VeritasAgent | Four ground-truth sources | VeritasRecord (provenance-tagged) | Assembles unified patient record from EHR, conversation logs, vital sign streams, and patient-reported input; assigns provenance tags to all fields |
| 2 | SentinelLayer | VeritasRecord | CandidateAlert | Detects parameter anomalies exceeding configured |

| | | | | thresholds; emits alert only when provenance-tagged parameters are available |
|---|---|---|---|---|
| 3 | DirectorAgent | CandidateAlert | Routed alert(s) | Routes CandidateAlert to relevant specialist agent(s) based on alert type classification and provenance tag profile |
| 4 | Specialist Agents (x6) | Routed alert | AgentClaim (risk level + recommendation) | Domain-specific evaluation of alert within clinical context; returns suppression or escalation recommendation with confidence |
| 5 | MetaSentinelAgent | All AgentClaims | SystemDecision | Aggregates specialist claims; applies cooldown/debounce logic; resolves inter-agent conflicts; routes final decision to DashboardService |

## 3.2 Specialist agent domains

Six specialist agents handle distinct false positive categories:

**ProbeIntegrityAgent:** Evaluates signal quality flags, probe cover status, and device connectivity to identify artefacts attributable to probe displacement or poor contact.

**ActivityIntegrityAgent:** Assesses accelerometer data and patient-reported activity state to contextualise motion artefact-induced SpO2 and HR anomalies.

**TachycardiaAgent:** Evaluates elevated HR readings against activity level, position, and recent clinical history to distinguish physiological tachycardia from artefact or baseline variation.

**BradycardiaAgent:** Assesses low HR readings in context of medication history, nocturnal state, and patient baseline to suppress medication-induced or physiologically normal bradycardia.

**COPDAgent:** Evaluates SpO2 readings in COPD patients against patient-specific baseline hypoxaemia thresholds, preventing false escalation of chronic low saturation as acute deterioration.

**NocturnalAgent:** Contextualises nocturnal physiological patterns, positional SpO2 dip, sleep-related bradycardia, against time-of-day and activity state to suppress false nocturnal escalations.

## 3.3 Provenance-guided routing

A key design feature of Veritas-RPM is the use of provenance tagging. Each parameter in the VeritasRecord is assigned a provenance tag at ingestion: device-verified (sensor reading with quality flag), patient-reported (self-reported activity or symptom), EHR-derived (clinical baseline or medication history), or inferred (LLM-generated value). The DirectorAgent's routing logic and each specialist agent's suppression logic operate on provenance-tagged inputs, structurally preventing LLM-inferred values from propagating into alert suppression decisions without human confirmation. This design ensures that a hallucinated or miscalibrated inference cannot cause a true clinical event to be suppressed.

In the proprietary implementation used in this evaluation, all specialist agents employ deterministic, guardrail-bounded rule evaluation: clinical threshold bounds are encoded per domain (for example, $SpO_2 \geq 86\%$ is within acceptable range for a documented COPD patient, whereas $SpO_2 < 94\%$ triggers evaluation for a patient without chronic hypoxaemia). The clinical logic constitutes proprietary implementation detail and is not present in the published reference implementation, which contains TODO stubs in place of clinical rules, consistent with the IP-protection design of the architecture. This separation between the evaluable proprietary system and the open reference framework is intentional and is consistent with the pluggable clinical logic design principle described in Section 3.3.

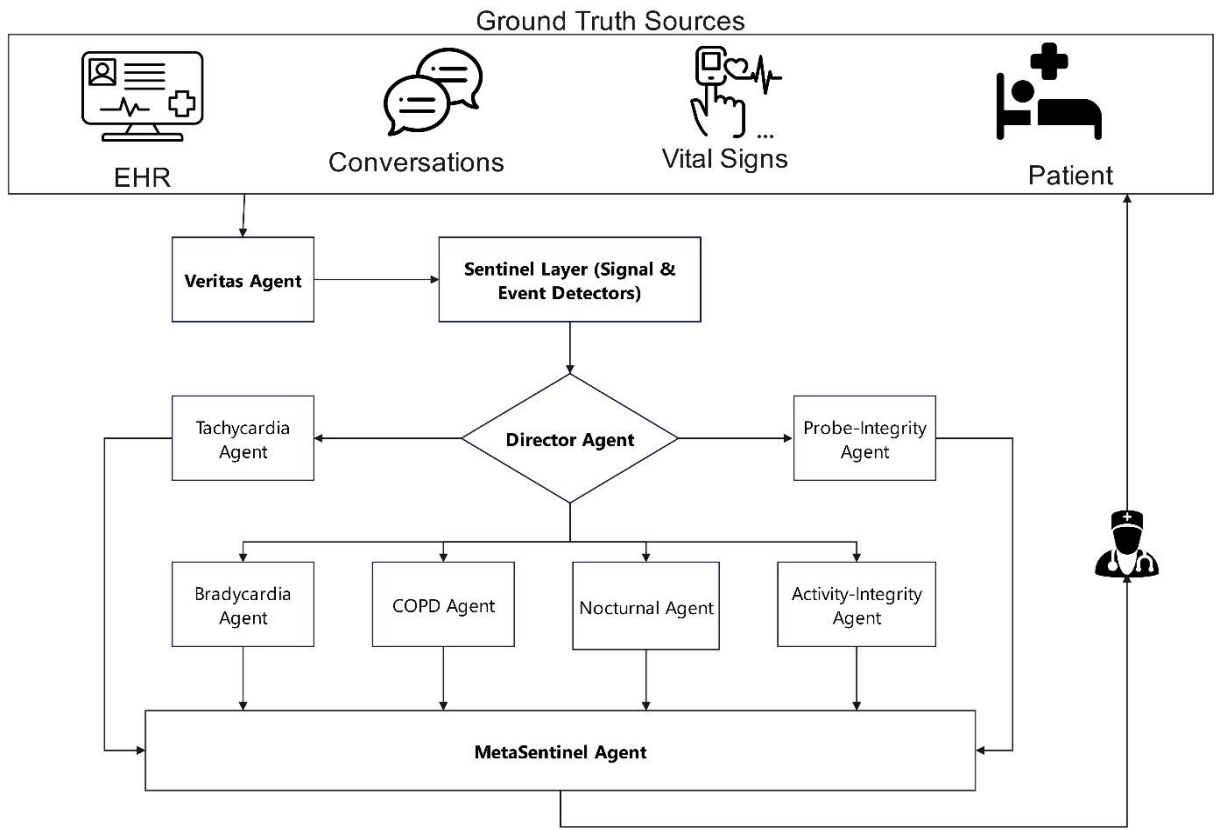

*Figure 1. Veritas-RPM five-layer provenance-guided multi-agent architecture.*
*Inputs from EHR, patient conversations, vital signs, and patient-reported data are assembled by the VeritasAgent with provenance tagging. The SentinelLayer detects anomalies and emits candidate alerts, which the DirectorAgent routes to one or more of six domain-specific Specialist Agents (TachycardiaAgent, ProbeIntegrityAgent, BradycardiaAgent, COPDAgent, NocturnalAgent, ActivityIntegrityAgent). The MetaSentinelAgent aggregates Specialist Agent claims and issues the final suppression or escalation decision to the clinical team.*

Specialist agent reasoning is conducted via structured prompts supplied with provenance-tagged parameters. The provenance layer ensures LLM-inferred values are structurally excluded from agent inputs; all parameters reaching specialist agents carry a device-verified, patient-reported, or EHR-derived tag. Temperature settings and model family used in the current simulation are specified in the Zenodo code repository.

Table 2. Implementation characteristics and determinism profile of each pipeline component.

| Component | Logic type | Deterministic? |
|---|---|---|
| VeritasAgent | Rule-based provenance tagger | Yes |
| SentinelLayer | Threshold-based signal detector | Yes |
| DirectorAgent | Rule-based classifier and router | Yes |
| Specialist Agents (×6) | Deterministic guardrail-bounded rule evaluation using provenance-tagged clinical parameters; clinical logic is proprietary and not present in the published reference implementation | Yes: fully deterministic within defined clinical bounds |
| MetaSentinel Agent | Weighted aggregation with deterministic conflict resolution rules | Yes: deterministic weighted aggregation; all specialist claims are rule-generated |

## 4. Methods

### 4.1 Study design

This study presents a simulation-based validation of the Veritas-RPM architecture consistent with pre-clinical evaluation methodology for medical device software. Empirically-informed synthetic time-series data were generated from a clinically grounded false-positive taxonomy. The taxonomy was derived through cross-source pattern analysis incorporating real patient RPM signal observations; no individual patient records, personal identifiers, or raw clinical data were transferred to or used by Teams B or C. The final simulation dataset is entirely synthetic in composition. The study does not constitute a clinical trial and does not require ethics committee approval under current research governance frameworks.

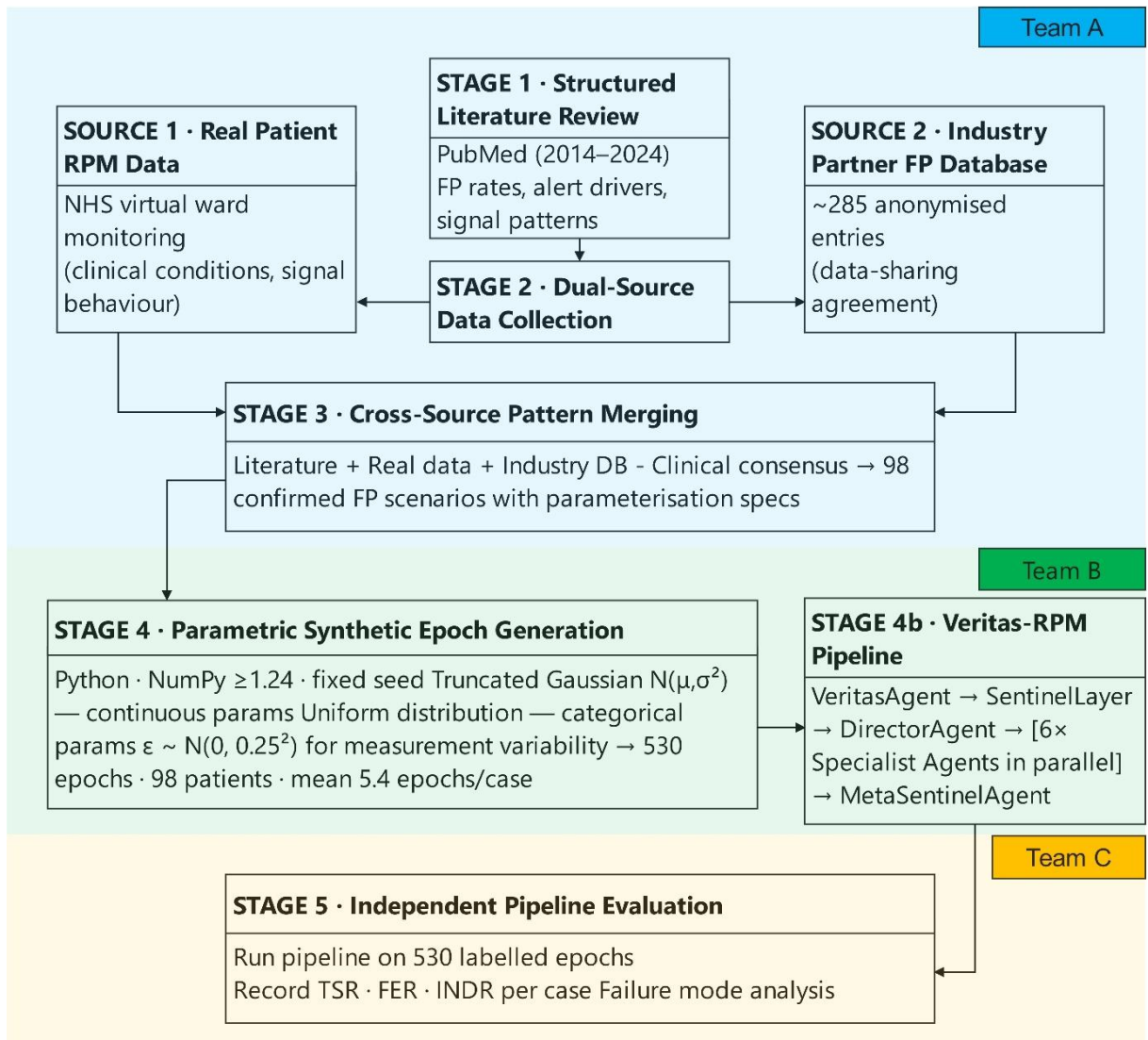

*Figure 2. Veritas-RPM data preparation and evaluation pipeline. Five stages across three operationally independent teams. Team A conducted all patient data handling and taxonomy construction; Team B received parameter specifications only and had no access to patient data; Team C evaluated system performance independently of both development and data collection.*

### 4.2 False positive taxonomy construction and pattern analysis (Phase 1)

> *Data governance statement:* Real patient RPM monitoring data reviewed by Team A during Stage 2 was obtained from monitoring records associated with NHS-partnered remote care services. Use of these records for pattern characterisation was conducted under data-sharing and clinical governance arrangements applicable to the respective service context. No individual patient data was transferred to, or accessed by, Teams B or C. The anonymised false positive reference database provided by the industry partner was used solely for pattern identification during taxonomy construction. No patient-identifiable information was received or processed. Use of this database was governed by a data-sharing agreement between the parties.

The 98-entry false positive taxonomy was constructed by Team A (clinical lead and medical informatician) through a three-stage process: structured literature review, dual-source data collection, and cross-source pattern identification and merging. Team A operated independently of Team B (architecture and simulation) and Team C (independent evaluation) throughout this phase.

*Stage 1: Structured literature review*

A structured review of peer-reviewed literature on false positive alert patterns in remote patient monitoring was conducted, covering publications from 2014 to 2024. The review was conducted across MEDLINE, Embase, and PubMed, with searches targeting false positive alert frequency, causes of misdetection, and signal-level characteristics associated with non-actionable alerts in continuous physiological monitoring systems. Although a substantial body of the identified literature concerns implantable loop recorder (ILR) remote monitoring, a cardiac implantable device context distinct from wearable pulse oximetry; these studies were included because they provide the most rigorous published quantification of false positive burden and alert driver taxonomy in remote monitoring, and their documented alert mechanisms (artefact, signal quality degradation, threshold-marginal values) are directly analogous to those encountered in PPG-based wearable systems.

Key findings informing the taxonomy included: (i) a systematic review of 3,305 ILR patients in which false positive alerts comprised 59.7% of all remote transmissions, with atrial fibrillation misdetection driven by premature atrial and ventricular complexes identified as the most common false positive alert type [Covino & Russo, 2024]; (ii) false positive transmission rates ranging from 46% to 86% depending on implant indication and device settings, with low R-wave amplitude and low-frequency filter configuration identified as significant device-related predictors [Afzal et al., 2019; Russo et al., 2023]; (iii) bradycardia false alerts attributable to undersensing of low-amplitude premature ventricular contractions, reducible through programming optimisation [Guarracini et al., 2022]; (iv) a remote monitoring alert burden study of 26,713 patients demonstrating that 54.8% transmitted at least one alert annually, with ILRs disproportionately contributing to non-actionable transmissions [O'Shea et al., 2020]; and (v) PPG-based wearable monitoring literature documenting motion artefact and probe displacement as systematic drivers of $SpO_2$ and HR false positives in community monitoring contexts [Knight et al., 2022].

*Stage 2: Dual-source data collection*

Team A collected and reviewed two complementary empirical sources: (i) a de-identified dataset of real patient RPM monitoring records obtained from NHS-partnered remote care services under applicable clinical governance arrangements, the size of which is subject to data-sharing confidentiality terms; this dataset provided empirical signal behaviour under clinical conditions representative of the NHS virtual ward context; and (ii) an anonymised false positive reference database of approximately 285 documented case entries provided by an industry partner under a data-sharing agreement.

*Stage 3: Cross-source pattern identification and merging*

The literature-derived false positive pattern catalogue and the two empirical data sources were analysed jointly to identify recurring signal signatures corroborated across all three inputs. Where a signal pattern was confirmed in the literature, present in the real patient data, and represented in the industry false positive database, it was retained as a confirmed taxonomy entry. Patterns present in only one source were reviewed by clinical consensus within Team A before inclusion. The cross-source merging process yielded 98 confirmed false positive scenarios spanning all six specialist agent domains. Each entry encodes the signal pattern, underlying physiological or technical cause, responsible agent domain, contextual reason for false positive classification, and the synthetic parameterisation specification subsequently used by Team B for epoch generation.

## 4.3 Synthetic data generation (Phase 2)

Synthetic patient epochs were generated by Team B using a parametric rule-based approach implemented in Python (NumPy ≥1.24). The parameter specifications for each taxonomy entry, including signal ranges, contextual flags, and episode characteristics, were received from Team A as a fixed, pre-specified input derived from the pattern-merging process described in §4.2. Team B had no role in defining these parameters.

For each taxonomy case, epoch parameter values were sampled from truncated Gaussian distributions $N(\mu, \sigma^2)$ bounded to physiologically plausible ranges $[a, b]$, with $\mu$ and bounds set directly from the confirmed pattern parameters. Uniform distributions were applied to categorical parameters (probe cover status, device status flag, patient position) and to threshold-marginal cases where a flat distribution across a boundary range was clinically appropriate. Minor stochastic noise $\epsilon \sim N(0, 0.25^2)$ was added to continuous parameters to reflect natural measurement variability in wearable pulse oximetry. A fixed random seed was applied to ensure full reproducibility; the generation script is available in the Zenodo repository (DOI: 10.5281/zenodo.19136793).

Across 98 patients, this produced 530 individual one-minute epoch rows (mean 5.4 epochs per case). Patient identifiers (3847291–3847388) and monitoring dates (June–August 2022) were assigned synthetically and carry no correspondence to real individuals.

### 4.4 Evaluation framework

Each synthetic epoch is derived from a taxonomy entry that was pre-classified as a confirmed false positive during Phase 1 pattern analysis. Accordingly, the correct output for every case, suppression, is known prior to pipeline execution, enabling objective measurement of system performance against a complete and unambiguous reference standard. This is the principal methodological advantage of simulation-based pre-clinical validation: controlled ground-truth availability that cannot be assumed in prospective real-world data. Agent performance is evaluated against three mutually exclusive outcome categories:

**True Suppression (TS):** System correctly identifies and suppresses a false positive alert.

**False Escalation (FE):** System incorrectly escalates a false positive to clinical staff.

**Indeterminate (IND):** System flags the alert for human review without escalation or suppression.

Primary metrics are the True Suppression Rate (TSR = TS/98), False Escalation Rate (FER = FE/98), and Indeterminate Rate (INDR = IND/98), reported overall and stratified by agent domain.

The unit of outcome evaluation is the case (n=98), not the epoch. Each case comprises one or more one-minute epochs representing a continuous monitoring window. Pipeline decisions are aggregated at the case level: a case is classified as True Suppression if all epochs in the case receive a suppression decision; as False Escalation if any epoch receives an escalation decision; and as Indeterminate if any epoch remains unresolved and no epoch triggers escalation. Domain performance figures in Table 2 are reported at case level using this aggregation rule.

## 5. Results

### 5.1 Overall performance

Across 98 cases and 530 epochs, the Veritas-RPM pipeline achieved an overall True Suppression Rate of 83.7% (82/98 cases) and a False Escalation Rate of 16.3% (16/98 cases). The Indeterminate Rate was 0% - every case received a definitive suppression or escalation decision. No case remained unresolved. This absence of indeterminate outcomes reflects the MetaSentinelAgent's mandatory final resolution step, which forces a binary decision even under inter-agent conflict, albeit at the cost of reduced accuracy in conflict scenarios.

Table 3. Overall pipeline performance across 98 simulated false positive cases (530 epochs).

| Metric | Count | Rate | Notes |
| --- | --- | --- | --- |
| True Suppression (TS) | 82 / 98 | 83.7% | System correctly suppresses false positive |
| False Escalation (FE) | 16 / 98 | 16.3% | System incorrectly escalates to clinical staff |
| Indeterminate (IND) | 0 / 98 | 0.0% | No unresolved cases - all received final decision |
| Total cases | 98 | - | 530 total 1-minute epochs, mean 5.4 per case |

## 5.2 Performance by agent domain

Five of the six specialist agent domains achieved 100% TSR across all evaluated cases. ProbeIntegrityAgent (n=23), ActivityIntegrityAgent (n=8), COPDAgent (n=13), BradycardiaAgent (n=2), and NocturnalAgent (n=3) each suppressed all assigned false positive cases without exception. TachycardiaAgent achieved 88% TSR (n=8), with one false escalation attributable to an isolated high HR reading without sufficient contextual signal to confirm artefact origin. Confidence intervals for domains with small sample sizes are wide and should be interpreted cautiously: BradycardiaAgent (n=2, 95% CI 34–100%) and NocturnalAgent (n=3, 95% CI 44–100%) achieved perfect suppression in all evaluated cases, but the sample sizes are insufficient to support broad performance claims for those domains.

Failures concentrated in two structural categories: MetaSentinelAgent conflict resolution cases (TSR 70%, FER 30%, n=30) and ProbeIntegrity combined with condition-specific cases (TSR 0–62%, n=11). These patterns are analysed in detail in Section 5.3.

Table 4. True Suppression Rate and False Escalation Rate stratified by specialist agent domain (n = 98 cases).

| Agent domain | n | TS | FE | TSR | FER |
| --- | --- | --- | --- | --- | --- |
| ProbeIntegrityAgent | 23 | 23 | 0 | 100% | 0% |
| ActivityIntegrityAgent | 8 | 8 | 0 | 100% | 0% |
| COPDAgent | 13 | 13 | 0 | 100% | 0% |
| BradycardiaAgent | 2 | 2 | 0 | 100% | 0% |
| NocturnalAgent | 3 | 3 | 0 | 100% | 0% |
| TachycardiaAgent | 8 | 7 | 1 | 88% | 12% |
| MetaSentinelAgent (conflict resolution) | 30 | 21 | 9 | 70% | 30% |
| ProbeIntegrity + Activity conflict | 8 | 5 | 3 | 62% | 38% |
| ProbeIntegrity + condition conflict | 3 | 0 | 3 | 0% | 100% |
| Total | 98 | 82 | 16 | 83.7% | 16.3% |

Table 5. Wilson 95% confidence intervals for TSR in specialist agent domains with limited case volumes.

| Agent domain | n | TSR | 95% CI |
|---|---|---|---|
| BradycardiaAgent | 2 | 100% | 34.2–100% |
| NocturnalAgent | 3 | 100% | 43.9–100% |
| ProbeIntegrityAgent | 23 | 100% | 85.2–100% |
| COPDAgent | 13 | 100% | 77.2–100% |

For domains with smaller sample sizes, 95% Wilson confidence intervals are reported separately to contextualise the precision of TSR estimates (Table 5).

### 5.3 Failure mode analysis

All 16 false escalations were examined for device status and alert type at the time of failure. Table 6 summarises the distribution of failure modes by device status flag.

Table 6. Distribution of false escalation failure modes by device status flag at time of failure (n = 16 cases).

| Device status at failure | Count | Primary failure mechanism |
|---|---|---|
| system_flag | 7 | MetaSentinelAgent unable to resolve competing specialist claims when system-level flag is present without corroborating provenance context |
| ok (no artefact flag) | 4 | Multi-signal conflict: ProbeIntegrity + condition agents in genuine disagreement on borderline SpO2 values (88.0–97.0%) |
| motion_artefact | 2 | ProbeIntegrity + Activity conflict: dual-agent routing with insufficient provenance resolution at MetaSentinel |
| probe_cover | 1 | ProbeIntegrity + Activity routing conflict; probe cover flag present but activity context insufficient |
| threshold_marginal | 1 | MetaSentinelAgent: SpO2 93.5%, HR 101.8 bpm - values at threshold boundary preventing definitive specialist claim |
| duplicate_alert | 1 | MetaSentinelAgent: duplicate alert pattern not resolved by cooldown logic in current implementation |

The pattern is interpretable: false escalations are not distributed randomly across the dataset but cluster in scenarios where (a) the MetaSentinelAgent receives conflicting specialist claims without a dominant provenance signal to resolve the conflict, or (b) parameter values fall at threshold boundaries where neither suppression nor escalation has sufficient evidence to reach the configured confidence threshold. In seven of sixteen cases, the device status flag was system_flag - a state indicating a platform-level alert without specific artefact classification - which provides insufficient provenance context for specialist routing and forces MetaSentinel resolution without adequate input.

Critically, no false escalations were produced in cases where a single specialist agent had clear domain ownership of the alert and full provenance context was available. All single-domain agent false escalation rates are 0% except TachycardiaAgent (12.5%, n=1), where the failure involved an isolated elevated HR without co-occurring provenance signals.

## 6. Discussion

### 6.1 Interpretation of results

The overall TSR of 83.7% is meaningful in the context of NHS RPM deployment. Current platform architectures produce false positive alert rates of 74–99%, meaning the overwhelming majority of alerts generated by threshold-based systems are non-actionable. A suppression layer that correctly resolves 83.7% of false positive cases before alerts reach clinical staff, with zero indeterminate outcomes, represents a substantial reduction in clinical alert burden. The 0% indeterminate rate is architecturally significant: the MetaSentinelAgent's design enforces a final decision on every case, avoiding the "system could not decide" outcome that would place unresolved cognitive load back on clinical staff.

The failure pattern strengthens rather than undermines confidence in the architecture. All false escalations originate from a single structural source: cases where two or more specialist agents receive overlapping domain claims and the MetaSentinelAgent lacks sufficient provenance context to resolve the conflict with confidence. This suggests the suppression logic is functionally sound in single-domain simulated scenarios; the MetaSentinel's conflict resolution weighting requires refinement for multi-signal cases. This is a precisely targeted and tractable engineering problem - not a fundamental limitation of the provenance-guided approach.

### 6.2 The system_flag failure cluster

Seven of sixteen false escalations involved device_status = system_flag, a platform-generated flag that indicates a system-level alert without specific artefact classification. In the current implementation, system_flag does not carry sufficient provenance granularity for the DirectorAgent to assign specialist routing with confidence, and the MetaSentinelAgent's conflict resolution logic defaults to escalation under ambiguity - a conservative safety design decision. Addressing this cluster requires either richer provenance tagging of system_flag states in the VeritasAgent layer, or a dedicated SystemFlagAgent that handles platform-level alerts before specialist routing. Either approach is architecturally straightforward and represents the primary engineering target for Phase 1 development.

### 6.3 Simulation methodology and external validity

The use of synthetic patient data parameterised from documented false positive patterns is an acknowledged limitation and an explicit methodological choice. Simulation enables controlled evaluation against known ground-truth labels, which is not achievable with prospective real-world data in early-stage development. The false positive taxonomy (98 cases) was constructed from documented case patterns consistent with published RPM literature and current NHS platform deployments. The simulated input model - SpO2, HR, accelerometer, device status flags, and patient-reported activity state - is consistent with data already collected by deployed NHS platforms including Doccla, Huma, Luscii/Graphnet, and Whzan Digital Health.

Synthetic data cannot fully capture real-world complexity: multi-parameter co-dependencies, patient-specific noise profiles, rare artefact combinations, and variable patient adherence to self-reporting are all absent from this dataset. The simulation supports the feasibility of the architecture under the conditions tested and that its failure modes are interpretable; it does not establish generalisability to real patient populations. Prospective validation on real-world RPM cases is the necessary and planned next step.

### 6.4 Phase 1 validation plan

Phase 1 prospective validation will evaluate Veritas-RPM on 200 curated real-world NHS RPM cases. Dataset construction follows a pre-defined quality protocol: records will be excluded for incomplete physiological streams, and all included cases will undergo dual-specialist conformity assessment and data integrity review prior to inclusion. The primary success criterion is a ≥40% reduction in false escalation rate compared to a single-LLM baseline agent (GPT-4o, no provenance tagging, no specialist routing) evaluated on the same dataset. Secondary criteria include MetaSentinelAgent conflict resolution TSR ≥80% on multi-signal cases and system decision latency ≤2 seconds per epoch.

The planned integration context for prospective validation is deployment downstream of an existing NHS RPM platform - preferably Doccla or Luscii/Graphnet, whose data models are most consistent with the simulated input parameters. This would enable real-world evaluation without requiring modification of the primary monitoring infrastructure.

## 7. Limitations

Because the synthetic cases were generated from a taxonomy derived from the same conceptual false-positive categories the architecture is designed to address, the present evaluation constitutes a preclinical test of internal architectural validity rather than an estimate of real-world deployment performance. The taxonomy-to-simulation pipeline by design eliminates distributional mismatch between test data and system assumptions, a condition that will not hold in prospective real-world evaluation. This circularity is an inherent property of simulation-based pre-clinical validation and is acknowledged as a primary limitation of the current phase.

Synthetic data cannot fully capture the complexity of real-world RPM streams, including multi-parameter co-dependencies, patient-specific noise profiles, or rare signal artefact combinations not represented in the 98-case taxonomy.

Ground truth labels are derived from documented case patterns rather than prospectively collected patient data, which may introduce classification bias in edge cases where alert categorisation is ambiguous.

The Ambient Condition parameter included in the epoch schema represents a future intended data source not yet standard in current UK RPM deployments. Results involving this parameter should be interpreted in the context of planned rather than currently available infrastructure.

Patient self-report context flags (activity level, position) are assumed accurate in the synthetic dataset. In practice, adherence to self-reporting in elderly or cognitively impaired cohorts, populations with the highest RPM monitoring burden, may be substantially lower, affecting ActivityIntegrityAgent and NocturnalAgent performance.

This study does not include a comparative baseline - performance is reported for Veritas-RPM in isolation against known labels. A prospective comparative evaluation against threshold-based alerting and single-LLM baseline approaches is required before quantitative superiority claims can be made.

## 8. Ethical Considerations

All data used in this study are synthetic. No real patient data, personal identifiers, or clinical records were used: all epoch values were computationally generated, and no individual patient data was directly processed by the pipeline or evaluation framework. Real patient RPM monitoring records were accessed exclusively by Team A during taxonomy construction, under data-sharing and clinical governance arrangements applicable to the respective NHS-partnered service context, for the purpose of pattern characterisation only. No patient identifiers, raw waveforms, or individual clinical records were transferred to Teams B or C, incorporated into the synthetic dataset, or used in pipeline evaluation. The anonymised false positive reference database provided by the industry partner contained no patient-identifiable information. This study does not constitute a clinical trial and does not require ethics committee approval under current research governance frameworks. Any future prospective validation involving NHS trust RPM infrastructure should seek IRB approval and NHS DTAC compliance review prior to patient data access.

## 9. Conclusion

Veritas-RPM indicates that provenance-guided multi-agent routing can achieve high false positive suppression across single-domain RPM alert types in simulation. The architecture's five-layer design - VeritasAgent provenance assembly, SentinelLayer anomaly detection, DirectorAgent specialist routing, six domain Specialist Agents, and MetaSentinelAgent conflict resolution - produces interpretable, domain-localised failure modes that directly inform the engineering targets for Phase 1 real-world validation.

The key finding - 100% TSR across five specialist agent domains with failures concentrated exclusively in MetaSentinelAgent conflict resolution for multi-signal cases - suggests the suppression logic is functionally sound under the simulated conditions tested. The MetaSentinel conflict resolution mechanism, particularly under system_flag device states and multi-domain signal overlap, represents the primary development target. Addressing this through richer provenance tagging or a dedicated system-flag handler is a candidate refinement for Phase 1 evaluation.

A Python reference implementation is available at https://zenodo.org/records/19136793 and https://github.com/justbetter21/veritas-rpm under MIT licence. Phase 1 prospective validation on 200 curated NHS RPM cases with dual-specialist conformity assessment is planned as the next stage of development.

## Author Contributions

| Author | Team | Specific contributions |
| --- | --- | --- |
| Aswini Misro (First author) | Team A - Clinical lead & medical informatician | Conceptualisation and specification of all six specialist agent domains; design of physiological thresholds for $SpO_2$, HR, and activity-state parameters; structured literature review (2014–2024, MEDLINE/Embase/PubMed) across all agent domains; collection and curation of real patient RPM monitoring data under applicable governance arrangements; collection and curation of the 285-entry false positive reference database; cross-source pattern identification and merging; construction of the final 98-case false positive taxonomy; conformity assessment of all 98 synthetic cases against clinical ground truth; differentiation of pathological from physiological signal changes across all six agent |
| | | domains; clinical validation of simulation parameter ranges; manuscript review and approval. |
| Vikash Sharma (Second author) | Team B - Chief data scientist & software architect | Design and implementation of the five-layer Veritas-RPM software architecture (VeritasAgent, SentinelLayer, DirectorAgent, Specialist Agents, MetaSentinelAgent); provenance-tagging mechanism and specialist agent routing logic; translation of Team A's pattern specifications into NumPy-based parametric sampling distributions; synthetic epoch generation (Python, NumPy ≥1.24); pipeline execution and results extraction; Python reference implementation (Zenodo DOI: 10.5281/zenodo.19136793); manuscript drafting and revision. |
| Shreyank Narayana Gowda (Third author) | Team C - Independent evaluator & supervisor | Affiliation: School of Computer Science, University of Nottingham, Nottingham, UK. Email: shreyank.narayanagowda@nottingham.ac.uk. Contributions: Independent oversight and supervision of simulation methodology; evaluation framework design (TSR/FER/INDR metrics); machine learning methodology review; manuscript review and approval. |

## Acknowledgements

The authors acknowledge the contribution of a group of research interns from multiple UK universities who assisted Team A with structured literature screening and data collation tasks as part of their supervised research placements. Their work contributed to the breadth of the literature review and the initial assembly of the false positive pattern catalogue. Individual names are not listed at their request. The industry partner who provided the anonymised false positive reference database under data-sharing agreement is acknowledged for their support of this work. The authors also acknowledge YouDiagnose Limited for providing the research and development environment in which this work was conducted.

## Conflict of Interest and Funding Statement

Aswini Misro and Vikash Sharma are affiliated with YouDiagnose Limited, the organisation in which this research was conducted. No external funding was received for this study. The industry partner who provided the anonymised false positive reference database had no role in study design, analysis, or manuscript preparation. Shreyank Narayana Gowda declares no conflicts of interest. The authors confirm that YouDiagnose Limited's commercial interests in remote patient monitoring technology represent a potential conflict of interest, which is disclosed here in the interests of transparency.